\documentclass[aps,prb,twocolumn,showpacs,superscriptaddress,10p,longbibliography]{revtex4-1}

\usepackage{amsmath}
\usepackage{amssymb}
\usepackage[colorlinks=true,linkcolor=red]{hyperref}
\usepackage[sort&compress]{natbib}
\usepackage{graphicx}
\usepackage{color}

\begin{document}

\title{Resonant frequencies and spatial correlations in frustrated arrays of Josephson type nonlinear oscillators}

\author{A.  Andreanov}
\affiliation{Center for Theoretical Physics of Complex Systems, Institute for Basic Science (IBS), Daejeon 34051, Republic of Korea}

\author{M. V. Fistul}
\affiliation{Center for Theoretical Physics of Complex Systems, Institute for Basic Science (IBS), Daejeon 34051, Republic of Korea}
\affiliation{Russian Quantum Center, National University of Science and Technology "MISIS", 119049 Moscow, Russia}

\date{\today}

\begin{abstract}
    We present a theoretical study of resonant frequencies and spatial correlations of Josephson phases in frustrated arrays of Josephson junctions. Two types of one-dimensional arrays, namely, the diamond and sawtooth chains, are discussed. For these arrays in the linear regime the Josephson phase dynamics is characterized by multiband dispersion relation $\omega(k)$, and the lowest band becomes completely \textit{flat} at a critical value of frustration, $f=f_c$ . In a strongly nonlinear regime such critical value of frustration determines the crossover from non-frustrated ($0<f<f_c$) to frustrated  ($f_c<f<1$) regimes. The crossover is characterized by the thermodynamic spatial correlation functions of phases on vertices, $\varphi_i$, i.e. $C_p(i-j)=\langle\cos[p(\varphi_i - \varphi_j)]\rangle$ displaying the transition from long- to short-range spatial correlations. We find that higher-order correlations functions, e.g. $p=2$ and $p=3$, restore the long-range behavior deeply in the frustrated regime, $f\simeq 1$. Monte-Carlo simulations of the thermodynamics of frustrated arrays of Josephson junctions are in good agreement with analytical results. 
\end{abstract}
%\pacs{42.50.-p,74.81.Fa,74.50.+r}

\maketitle

\section{Introduction}

Great attention has been devoted to  theoretical and experimental study of dynamics of various systems of interacting nonlinear oscillators. Interesting effects, e.g. collective (synchronized) behavior,~\cite{pikovsky2003synchronization,filatrella2000highq} electromagnetic field induced dynamic metastable states,~\cite{jung2014multistability} solitons and breathers,~\cite{ustinov1998solitons,binder2000observation} just to name a few, were predicted and observed in diverse solid state systems. 

In the linear regime the dynamics of complex networks of interacting oscillators is characterized by a multiband dispersion relation $\omega_m(k)$, where $k$ is the wave vector of the extended linear excitations and $m$ is the band index. The dispersion relation can be probed by a resonant response of a system to a small external time-dependent perturbation.~\cite{jung2014multistability,macklin2015near}

In Ref.~\onlinecite{lieb1989two} it was predicted that the spectrum of electronic excitations of the Lieb lattice contains a \emph{flatband}. Lately flatbands were identified theoretically in various one- and two-dimensional lattices~\cite{derzhko2007low,vidal2000interaction,douccot2002pairing,vidal2001disorder,leykam2013flat} and several methods to engineer flatbands were suggested.~\cite{flach2014detangling,dias2015origami,morales2016simple,maimaiti2017compact,ramachandran2017chiral,roentgen2017compact} Flat bands have been observed experimentally in magnetic and high-$T_c$ superconducting materials,~\cite{liu2013flat,dessau1993key} and they have been implemented in photon lattices,~\cite{vicencio2015observation} exciton-polariton condensates,~\cite{byrnes2014exciton} and arrays of superconducting Josephson junctions.~\cite{pop2008measurement}

Dynamics of electrons, photons or phonons on lattices supporting flatbands shows interesting physical properties: magnetic phase transitions,~\cite{derzhko2007low} localization in absence of disorder --- appearance of \emph{compactons}: eigenstates strictly localized on few sites, destructive interference of electromagnetic waves propagating in such lattices. In presence of disorder or non-linearity the flatband spectrum leads to solitons with non-exponential tails,~\cite{gorbach2005compactlike} Fano resonances in the scattering of  electromagnetic waves on nonlinearities,~\cite{flach2014detangling} and topological effects, e.g. preservation of topological flatbands in applied magnetic field.~\cite{aoki1996hofstadter} 

The case of Josephson junctions is of special interest since the current technology allows one to engineer arbitrary one- and two-dimensional lattices of coupled Josephson junctions. Furthermore,  external magnetic fields allow one to change the ground state and the dispersion relation $\omega(k)$ of linear  oscillations in these systems. E.g. it was shown theoretically that a simple diamond chain of identical Josephson junctions exhibits a classical (quantum) phase transition at specific strengths of the applied magnetic field.~\cite{douccot2002pairing,rizzi20064e,protopopov2004anomalous} This phenomenon was dubbed $4e$- condensation at variance with the usual $2e$-condensation occurring in simple Josephson junction arrays (lattices). Some evidences in support of such a $4e$-condensation have been reported in Ref.~\onlinecite{pop2008measurement}.

In this paper we present a systematic study of dynamic and thermodynamic properties of \emph{frustrated Josephson junction arrays}. In these periodic arrays the Josephson couplings in a single cell can have alternating signs giving rise to frustration, which is quantified by a frustration parameter $f$, and the properties of the arrays, namely, resonant frequencies and ground states depend strongly on  the value of $f$.  In particular, we find a crossover between non-frustrated and  frustrated regimes characterized by the critical value, $f=f_c$, and study in detail the frustrated regime, $f_c<f<1$. We show that the lowest band in the linear spectrum $\omega(k)$ turns flat at $f=f_c$, and this can be considered as the precursor of the crossover. The most spectacular difference between the non-frustrated and frustrated regimes is in the properties of the ground states. The ground state in non-frustrated regime is unique, with all Josephson phases equal to zero. In contrast, the ground state in frustrated regime is macroscopically degenerate and the Josephson phases can take two different sets of values in each cell of the array. Note that at variance with previous works in this field~\cite{douccot2002pairing,rizzi20064e,protopopov2004anomalous} where high degeneracy occurred at a single value of $f$, this highly degenerate ground state occurs in a range of frustration values, $f_c<f<1$.  

Experimental realizations of such arrays requires Josephson couplings of different signs. Such Josephson couplings are provided by the so-called  $\pi$-Josephson junctions that can be fabricated on basis of superconductor-ferromagnet-superconductor junctions,~\cite{feofanov2010implementation} or various multi-junctions SQUIDS in externally applied magnetic field.~\cite{hilgenkamp2008pi}

The paper is organized as follows: In Section II we introduce models of frustrated arrays of Josephson junctions:  the diamond  and the sawtooth chains. Next we define the Lagrangian, partition function and dynamic equations of the systems. In Section III we analyze the linearized dynamics of the frustrated arrays, derive the dispersion relation $\omega(k)$ and study its dependence on the frustration strength $f$. Section IV is devoted to the derivation of the thermodynamic spatial correlation functions as functions of temperature and frustration, and  detailed discussion of the crossover between frustrated and non-frustrated regimes. Section V presents numerical support for the analytical results of the previous sections by the Monte-Carlo simulations. Section VI provides conclusions.

%    Figure 1
\begin{figure}
    \includegraphics[width=0.95\columnwidth]{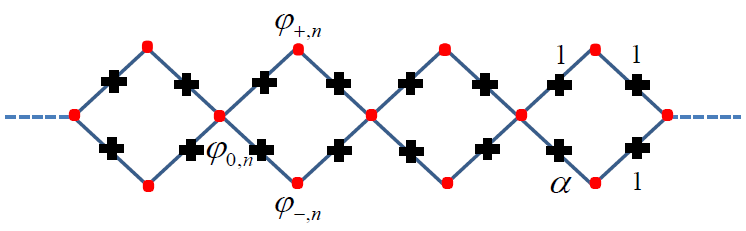}
    \includegraphics[width=0.95\columnwidth]{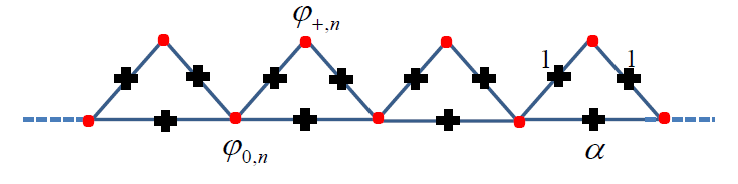}
    \caption{(Color online) Schematic figure of frustrated networks of Josephson junctions (indicated by crosses): the diamond (top) and  sawtooth (bottom) chains. The phases $\varphi$ of the vertices and the Josephson couplings $\alpha_{ij}$ in a single cell are shown.}
    \label{fig:schematic}
\end{figure}

\section{Model, Lagrangian, Partition function and Dynamic Equations}

The diamond and the sawtooth chains are shown in Fig.~\ref{fig:schematic}. Every vertex $i$ of the network is characterized by a time-dependent phase, $\varphi_i(t)$. Nearest neighbor vertices are binded by Josephson type non-linear oscillators, and therefore, the Lagrangian of network is as follows:
\begin{equation}
    \label{Lagrangian}
    L\{\varphi_i, \dot \varphi_i \} = E_J\left [\sum_i \frac{\dot \varphi_i^2}{2\omega_p^2} + \sum_{\langle ij\rangle} \alpha_{ij}\cos (\varphi_i - \varphi_j) \right ],
\end{equation}
where $\langle ij\rangle$ are the two nearest neighbor vertices coupled by a Josephson junction, $E_J$ and $\omega_p$ are the Josephson coupling strength and the plasma frequency, respectively. $\alpha_{ij}$ are the relative Josephson coupling strength of the $ij$-bond. The set of dynamic equations is then written as:
\begin{equation}
    \label{DynamicEq}
    \frac{E_J}{\omega_p^2}\ddot \varphi_i(t) = \frac{\partial L}{\partial \varphi_i}.
\end{equation}
Thermodynamic properties are given by the partition function $Z$ that can be expressed through the path integral in the imaginary time-representation:
\begin{equation}
    \label{PartFunction}
    Z = \int D[\varphi_n(\tau)] \exp \left [\frac{1}{\hbar}\int_0^{\hbar/(k_BT)}L\{\varphi_n, \dot \varphi_n, ~i\tau \} d \tau \right].
\end{equation}
The spatial correlations in the network of Josephson junctions are characterized by the correlation functions, $C_p(n)$, that are defined as:
\begin{gather}
    C_p(n) = \frac{1}{Z}\int D[\varphi_m(\tau)]\cos[p(\varphi_0 - \varphi_n)] \times\notag\\
    \times \exp \left [\frac{1}{\hbar}\int_0^{\hbar/(k_BT)}L\{\varphi_m, \dot \varphi_m, ~i\tau \} d \tau \right].
    \label{CorrFunction}
\end{gather}

Next we use the generic Eqs. (\ref{Lagrangian}--\ref{CorrFunction}) to analyze the dynamic and thermodynamic properties of the \emph{frustrated} arrays of coupled Josephson junctions. The frustrated arrays are characterized by specific distribution of Josephson coupling strengths, $\alpha_{ij}$, which display alternating signs of the couplings in every unit cell.

\section{Linear Regime: Dispersion relation $\omega(k)$}

\subsection{Diamond chain}

\begin{figure}
    \includegraphics[width=0.95\columnwidth]{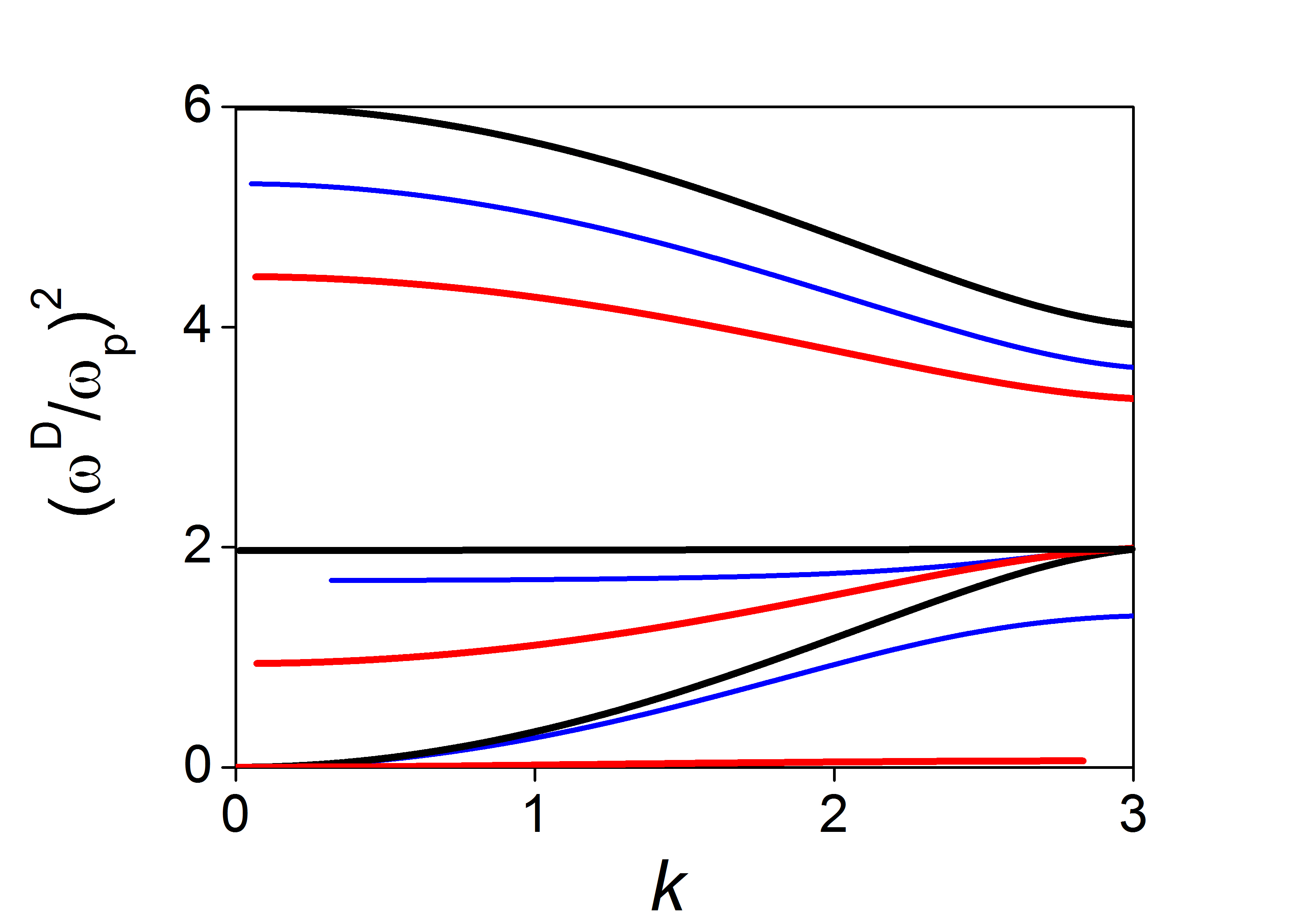}
    \caption{(Color online) The dispersion relation $\omega^D(k)$ for the diamond chain of Josephson junctions for several values of the frustration parameter: $f=0$ (black lines), $f=0.55$ (blue line), $f=0.65$ (red line). Flat bands occur for $f=0$ and $f=2/3$. The flatband at $f=2/3$ is the lowest one and marks the transition.}
    \label{fig:DChain}
\end{figure}

The state of the diamond chain of Josephson junctions is described by three phases per unit cell, $\varphi_n=\{\varphi_{0,n}, \varphi_{+,n}, \varphi_{-,n} \}$ (see the top part of Fig.~\ref{fig:schematic}. The distribution of Josephson coupling strength in a single cell is chosen as $\alpha_1=\alpha_2=\alpha_3=1$ and $\alpha_4=\alpha$. The parameter $\alpha$ can take values from $1$ to $-1$. For convenience, we define the frustration parameter $\alpha=1 - 2f$ varying from $0$ (non-frustrated arrays) to $1$ (maximal frustration). The precise choice of $\alpha_4$ is not important, and it can be swapped with any other $\alpha_i$. As we linearize Eq.~\eqref{DynamicEq} around the uniform solution, the dynamic equations become
\begin{align}
    \label{LinEquation-DC}
    \frac{1}{\omega_p^2}\ddot {\varphi}_{+,n} = & \,\,\varphi_{0,n} + \varphi_{0,(n+1)} - 2\varphi_{+,n} \;,\\
    \frac{1}{\omega_p^2}\ddot {\varphi}_{-,n} = & \,\,(1 - 2f)\varphi_{0,n} + \varphi_{0,(n+1)} - 2(1 - f)\varphi_{-,n} \;,\notag\\
    \frac{1}{\omega_p^2}\ddot {\varphi}_{0,n} = & \,\,\varphi_{+,n} + \varphi_{+,(n-1)} + (1-2f)\varphi_{-,n}\notag\\
    + & \varphi_{-,(n-1)} - 2(2 - f)\varphi_{0,n} \;.\notag
\end{align}
Using periodicity of the system and applying the Fourier transform with respect to space and time, we obtain the dispersion relation $\omega^D(k)$ as solutions of the following transcendent equation ($ \alpha = 1 - 2f$):
\begin{gather}
    -\frac{(\omega^D)^2}{\omega_p^2} + 3 + \alpha = \frac{4\cos^2 \left (\frac{k}{2} \right)}{2 - \omega^2} + \frac{(\alpha - 1)^2 + 4\alpha\cos^2 \left (\frac{k}{2} \right)}{-\omega^2 + 1 + \alpha}
    \label{TrEq-DC}
\end{gather}
This equation has three solutions $\omega_m^D(k)$, representing the three bands which are shown in Fig.~\ref{fig:DChain} for different values of the frustration parameter $f$. One can see that flatbands occur for two particular values of the frustration parameter, $f=0$ and $f=2/3$. In the latter case the flatband is \emph{the lowest band} of the spectrum, indicating the phase transition and the change of the ground state as the frustration parameter goes to the region $f>f^D_c=2/3$. 

%    Figure 2

\subsection{Sawtooth chain}

\begin{figure}
    \includegraphics[width=0.95\columnwidth]{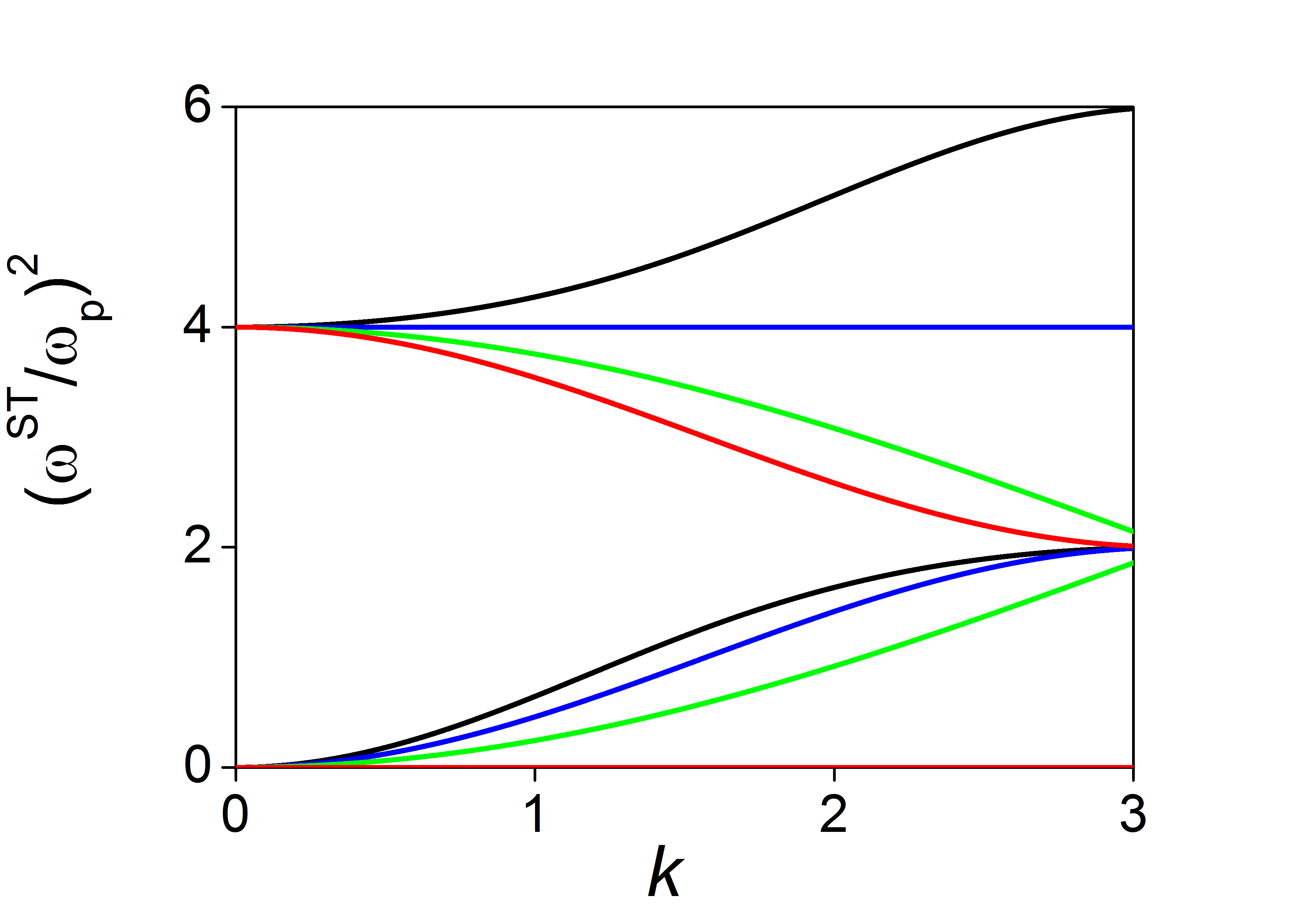}
    \caption{(Color online) The dispersion relation $\omega(k)$ for the sawtooth chain of Josephson junctions for several values of the frustration parameter: $f=0$ (black lines), $f=0.25$ (blue lines), $f=0.5$ (green lines),  $f=0.75$ (red lines). The flatbands occurring at $f=0.25$  and $f=0.75$ (the lowest one) are shown.}
    \label{fig:STchain}
\end{figure}

Similarly to the previous subsection, here, we analyze the dynamics of the sawtooth chain of Josephson oscillators (see the bootom part of Fig.~\ref{fig:schematic}). The sawtooth chain is described by two phases per unit cell, $\varphi_n= \{\varphi_{0,n}, \varphi_{+,n} \}$. The Josephson couplings strengths in a single cell (as shown in Fig.~\ref{fig:schematic}) is fixed to $\alpha_1=\alpha_2=1$ and $\alpha_3=\alpha=1-2f$. The frustration parameter $f$ varies from $0$ to $1$. The linearized dynamic equations~\eqref{DynamicEq} for the sawtooth chain are
\begin{align}
    \label{LinEquation-ST}
    \frac{1}{\omega_p^2}\ddot {\varphi}_{+,n} = & \,\,\varphi_{0,n} + \varphi_{0,(n+1)} - 2\varphi_{+,n} \;,\\
    \frac{1}{\omega_p^2}\ddot {\varphi}_{0,n} = & \,\,\varphi_{+,n} + \varphi_{+,(n-1)} + (1 - 2f)\varphi_{0,n-1}\notag\\
     + & (1 - 2f)\varphi_{0,(n+1)} - 2(1 - 2f)\varphi_{0,n} \;.\notag
\end{align}
The dispersion relation $\omega^{ST}(k)$ for time and space periodic solution reads 
\begin{equation}
    \label{Dispersion-Spectrum}
    \omega^{ST}(k) = \omega_p \left \{2 + 2\alpha\sin^2 \frac{k}{2} \pm \sqrt{4\alpha^2 \sin^4 \frac{k}{2} + 4\cos^2 \frac{k}{2}} \right \}^{1/2}
\end{equation}
There are two solutions for every $k$ corresponding to $2$ bands, shown in Fig.~\ref{fig:STchain} for several values of the frustration parameter $f$. The flatband occurs+ for two particular values of the frustration parameter, $f=0.25$ and $f=0.75$. Again, in the latter case the flatband is \emph{the lowest band} of the spectrum, and it marks the change of the ground state as the frustration parameter goes to the region $f>f^{ST}_c=3/4$. 
 
%    Figure 3

\section{Thermodynamics of frustrated arrays of Josephson junctions: spatial correlation functions}

In the classical regime the thermodynamic properties of the arrays of interacting Josephson junctions are described by the partition function
\begin{equation}
    \label{PF-DC}
    Z = \int D\{\varphi_i\}\exp \left[-\frac{U}{k_B T} \right ],
\end{equation}
where the potential energy $U\{\varphi_n\}$  depends on the set of vertex phases. The correlation functions  of interest are expressed as
\begin{equation}
    \label{Corrfunction-DC}
    C_p(n) = \frac{1}{Z}\int D\{\varphi_i\}\exp \left \{-\frac{U}{k_BT}+ip[\varphi_{0,m}-\varphi_{0,m+n}]\right \}
\end{equation}
Note that due to the global symmetry $\varphi_n\to-\varphi_n$ respected by the energy $U$ the above expression is real.

\subsection{Frustrated diamond chains}

In a particular case of diamond chain the potential energy is written as
\begin{gather}
    U\{\varphi_n\} = -E_J\sum_n \cos[\varphi_{+,n} - \varphi_{0,n}] + \cos[\varphi_{+,n} - \varphi_{0,n+1}]\notag\\   
    \label{Potenergy-DC}
    + \cos[\varphi_{-,n} - \varphi_{0,n+1}] + (1 - 2f)\cos[\varphi_{-,n} - \varphi_{0,n}]
\end{gather}
The correlation functions $C_p(n)$~\eqref{Corrfunction-DC} can be computed exactly for the chain by introducing new variables: $\varphi_{+,n}-\varphi_{0,n}=s_{1n}$, $\varphi_{0,n+1}-\varphi_{+,n}=s_{2n}$, $\varphi_{-,n}-\varphi_{0,n+1}=s_{3n}$ and $\varphi_{0,n}-\varphi_{-,n}=s_{4n}$. New variables satisfy a simple constraint, $ s_{1n}+s_{2n}+s_{3n}+s_{4n}=0$. Integrating over $s_{3n}$ and $s_{1n}-s_{2n}$ the spatial correlation functions evaluate to
\begin{gather}
    C^{D}_p(n) = \left \{ \frac{F^{D}_p}{F^{D}_0} \right \}^n\qquad\alpha = 1 - 2f,\notag\\
    \label{Corrfunction-DC-2}
    F^{D}_p = \int_0^{2\pi} du e^{ipu} I_0 \left [2K\cos \frac{u}{2} \right ]   I_0 \left [K\sqrt{1+\alpha^2+2\alpha \cos u} \right ],
\end{gather}
where $K=E_J/(k_B T)$ and $I_0(z)$ is the modified Bessel function.~\cite{abramowitz1964handbook} For high temperatures $k_B T\gg E_J$ ($K \ll 1$) one obtains the expected  strong suppression of correlations 
\begin{equation}
    \label{Corrfunction-DC-HT}
    C^D_p(n) \simeq  \left[\eta_p(f)\frac{E_J}{k_B T} \right]^{2pn},
\end{equation}
where $\eta_p(f)$ is a smooth function of order one that depends on both $p$ and $f$. For $p=1$ one has explicitly $\eta_1(f)=1-f$. 

The situation is less obvious in the low temperature regime as $k_B T\ll E_J$ ($K \gg 1$). The spatial decay of correlation functions strongly depends on the value of the frustration parameter $f$. Indeed, for $0<f<2/3$ the value $u \simeq 0$ gives a most important contribution to the integrals over $u$ in the Eq.~\eqref{Corrfunction-DC-2}, and therefore, the spatial correlation function displays long-range correlations as 
\begin{equation}
    \label{Corrfunction-DC-LT-Nofr}
    C^{D}_p(n) = \exp {\left [ -\frac{(1-f) k_BT}{(2-3f )E_J}p^2n \right ]}, ~~0<f<2/3
\end{equation}
Thus, one can see that in this regime the correlation length $\xi_p$ defined as $1/\xi_p=-\ln[C_p(n)]/n $, increases as $\xi_p\simeq 1/T$ with the decrease of temperature $T$. 

However in the highly frustrated regime as $f^D_c=2/3 < f < 1$, the values $u=\pm u_0=\pm 2\arccos [\sqrt{(f)/[2 (2f-1)]}]$ give the most important contribution to the integrals over $u$ in Eq.~\eqref{Corrfunction-DC-2}, and we obtain a crossover to a regime of short-range correlations as 
\begin{equation}
    \label{Corrfunction-DC-LT-fr}
    C^{D}_p(n) = \exp {\left [ -\frac{\beta_p(f) k_B T}{E_J}p^2n \right ]} \{ cos (pu_0) \}^n,
\end{equation}
where $\beta_p(f)$ is a smooth function of $p$ of order one.

\begin{figure}
    \includegraphics[width=0.95\columnwidth]{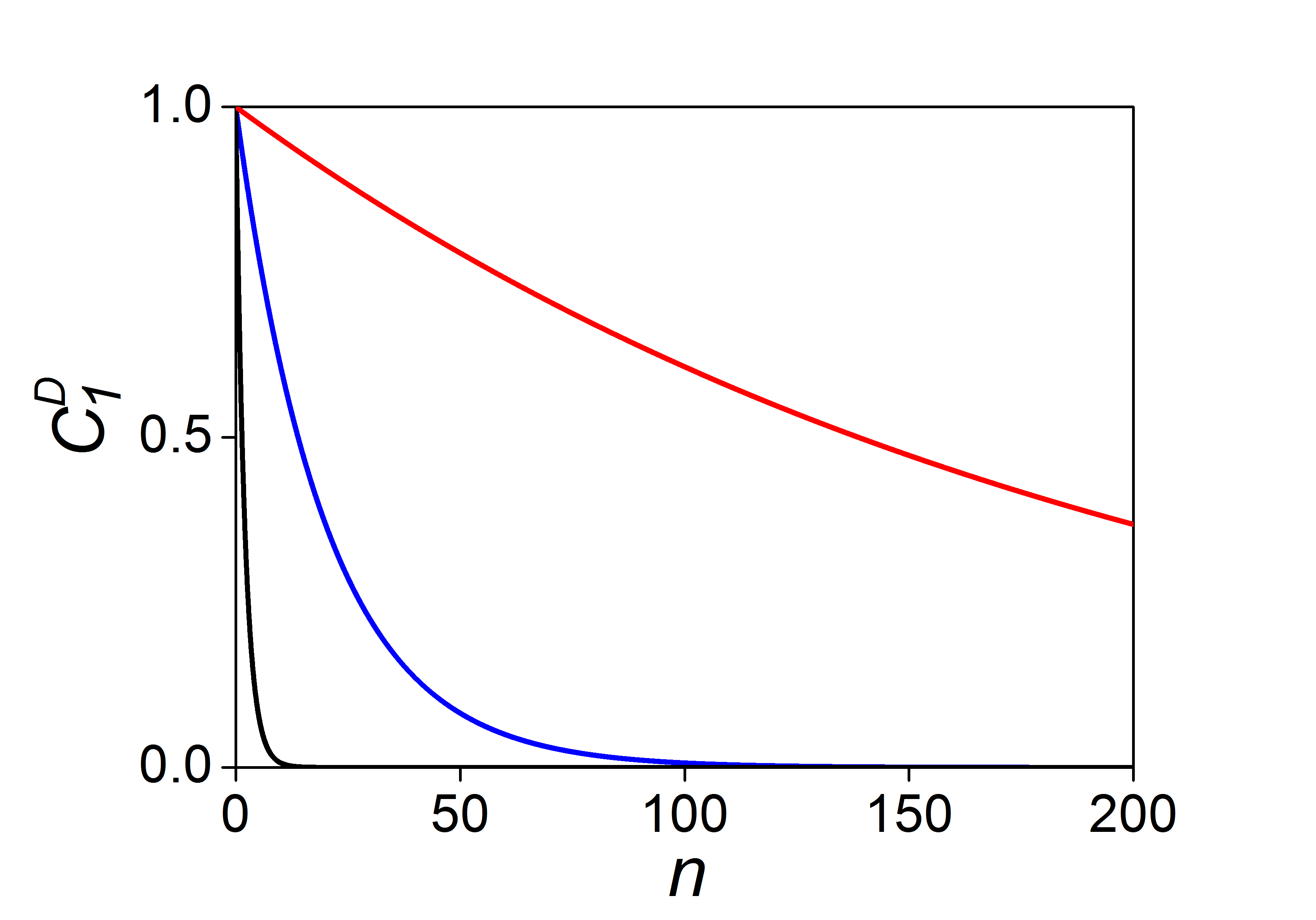}
    \includegraphics[width=0.95\columnwidth]{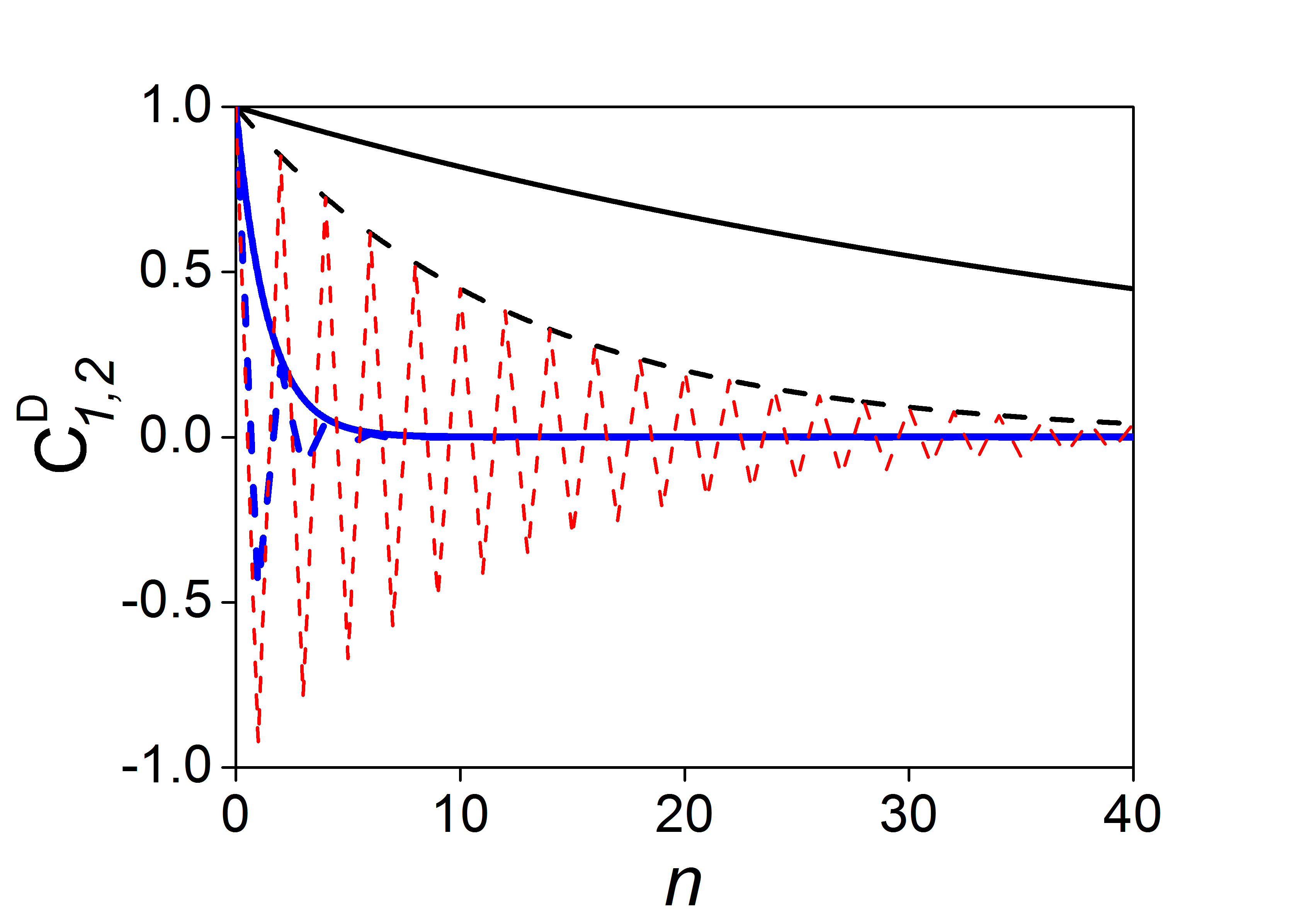}
    \caption{(Color online) The typical spatial dependencies of the diamond chain correlation functions, $C^{D}_1(n)$ (solid lines) and $C^{D}_2(n)$ (dashed lines), for different temperatures (top) and frustrations (bo+ttom). Top: the parameters $k_BT=E_J$ (black line)  $k_BT=0.1 E_J$ (blue line) and $k_B T=0.01 E_J$ (red line) and the frustration $f=0$ were chosen. Bottom: the frustration $f=0.5$ (black lines), $f=0.75$ (blue lines) and $f=1$ (red line), $k_BT=0.02 E_J$ were chosen.}
    \label{fig:Dchain-Cf}
\end{figure}

The correlation functions, $C^{D}_{p=1,2}(n)$, computed for several values of temperature and frustration are shown in Fig.~\ref{fig:Dchain-Cf}. As one can see the critical frustration $f^D_c=2/3$ that determines the crossover from long-range to short-range correlations, coincides with the value of $f$ where the lowest band in the spectrum $\omega(k)$ of linear excitations turns flat. This corresponds to a transition from unique ground state $u=0$, to a highly degenerate ground state characterized by two possible values of $u=\pm u_0$ that can be chosen independently for every unit cell. However, as we turn to higher order correlation functions with $p>1$ we find that e.g. for frustration $f \simeq 1$ the value of $u \simeq \pi/2$ and the correlation function $C_2(n)$ displays a long-range behavior with sign alternation on adjacent cells. This indicates the $2e$-$4e$ transition predicted and analyzed in Refs.~\onlinecite{douccot2002pairing,rizzi20064e,protopopov2004anomalous}. 

%    Figure 4

\subsection{Frustrated sawtooth chains} 

For the sawtooth chain of Josephson junctions the potential energy $U$ reads
\begin{gather}
    U\{\varphi_n\}=-E_J\sum_n \cos[\varphi_{+,n} - \varphi_{0,n}] + \cos[\varphi_{+,n} - \varphi_{0,n+1}]\notag\\
    \label{Potenergy-ST}
    + (1-2f)\cos[\varphi_{0,n} - \varphi_{0,n+1}]
\end{gather}
The correlation functions are computed similarly to the case of the diamond chain: the path integrals in the Eqs.~(\ref{PF-DC}--\ref{Corrfunction-DC}) are evaluated by introducing new variables $\varphi_{+,n}-\varphi_{0,n}=s_{1n}$, $\varphi_{0,n+1}-\varphi_{+,n}=s_{2n}$ and $\varphi_{0,n}-\varphi_{0,n+1}=s_{3n}$. The spatial correlation functions are expressed as ratios $C^{ST}_p(n)=(F^{ST}_p/F^{ST}_0)^n $, where 
\begin{equation}
    \label{Corrfunction-ST-2}
    F^{ST}_p = \int_0^{2\pi} du e^{ipu} I_0 \left [2K\cos \frac{u}{2} \right ]   e^{ -K(1-2f)\cos u },
\end{equation}
and $K$ and $I_0$ are the same as in the diamond chain case. For high temperatures $k_B T\gg E_J$ ($K \ll 1$) the correlation function $C^{ST}_1(n) \simeq [E_J/k_B T ]^{2n}$ shows fast exponential decay. In the low temperature regime as $k_B T\ll E_J$ ($K \gg 1$) the spatial decay of correlation functions strongly depends on frustration $f$ just like in the diamond chain case. Indeed, for $0<f<0.75$ the value $u \simeq 0$ gives the most important contribution to the integrals over $u$~\eqref{Corrfunction-ST-2}, and the spatial correlation function displays long-range correlations as
\begin{equation}
    \label{Corrfunction-ST-LT-Nofr}
    C^{ST}_p(n) = \exp {\left [ -\frac{ k_BT}{(3-4f) E_J}p^2n \right ]}, ~~f<0.75
\end{equation}
In the frustration regime $f^{ST}_c=0.75 <f< 1$ the values $u=\pm u_0=\pm 2\arccos [1/(4f-2)]$ give the most important contribution to the integrals over $u$ in~\eqref{Corrfunction-ST-2}, and we find a crossover to a regime of short-range correlations 
\begin{equation}
    \label{Corrfunction-ST-LT-fr}
    C^{ST}_p(n) = \exp {\left [ -\frac{ k_B T}{E_J}p^2n \right ]} \{ cos (pu_0) \}^n.
\end{equation}

\begin{figure}
    \includegraphics[width=0.95\columnwidth]{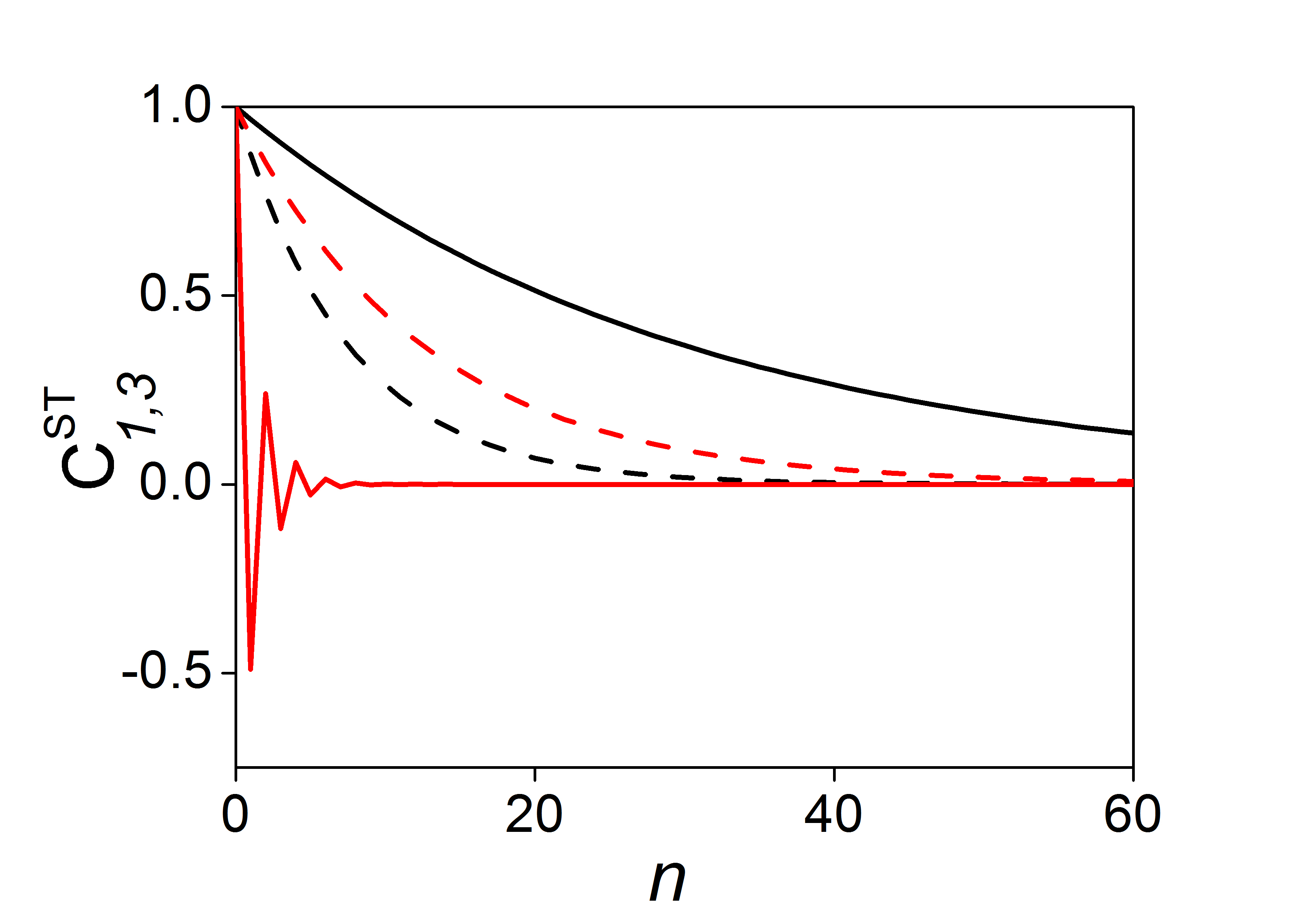}
    \caption{(Color online) The typical spatial dependencies of the sawtooth  chain correlation functions, $C^{ST}_1(n)$ (solid lines) and  $C^{TS}_3(n)$ (dashed lines) for different  frustrations:  $f=0.6$ (black line) and $f=1$ (red line).  The parameter $k_BT=0.02 E_J$ was chosen.}
    \label{fig:STchain-Cf}
\end{figure}

The correlation functions, $C^{ST}_{p=1,3}(n)$, computed for low temperature and for several  values of frustration are shown in Fig.~\ref{fig:STchain-Cf}. As for the diamond chain, the critical frustration $f^{ST}_c=3/4$ marks the crossover from long-range to short-range correlations and coincides with the value of $f$ where the lowest band in the spectrum $\omega(k)$ of linear excitations becomes flat. This indicates the transition from unique ground state $u=0$, to the highly degenerate ground state characterized by independent choice of any of the two values of $u=\pm u_0$ in every unit cell. However, as we turn to higher order correlation functions with $p>1$ we obtain that e.g. for $f \simeq 1$ the value of $u \simeq 2\pi/3$ and the correlation function $C_3(n)$ displays a long-range behavior. Notice here, that diamond chain  $C_2(n)$ displays oscillating behavior with $n$ at $f=1$ while the sawtooth model $C_3(n)$  does not oscillate with $n$. This difference --- sign alternation --- does not seem to be related to the difference in geometry of the two models: diamond chain is bipartite while sawtooth chain is not. One can modify the diamond chain model by adding a coupling between $\varphi_{n,+}$ and $\varphi_{n,-}$, that breaks the chiral symmetry of the Hamiltonian, nevertheless the $C_2(n)$ still shows oscillations as a function of $n$. 

%    Figure 5

\section{Spatial correlation functions: Monte-Carlo results}

In order to check the analytical predictions we performed classical Monte-Carlo simulations using the Hamiltonians~\eqref{Potenergy-DC} and~\eqref{Potenergy-ST}, treating the vertex phases $\varphi_{*,n}$ as classical XY spins $\vec{s}_{*,n}=(\cos\varphi_{*,n}, \sin\varphi_{*,n})$. We used systems with $N=900$ (diamond chain) and $N=600$ (sawtooth chain) sites, each corresponding to 300 unit cells. A typical run consisted of a simulated annealing  where the system was heated up from $k_B T = 0.01 E_J$ to $10 E_J$ in $100$ steps. A set of correlation functions $C_p(n)$ with $p=1\dots 8$ was computed during the annealing.

%    Figure 6
\begin{figure}
    \includegraphics[width=0.95\columnwidth]{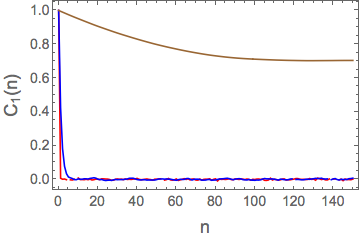}
    \includegraphics[width=0.95\columnwidth]{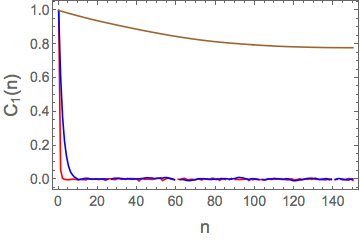}
    \caption{(Color online) The spatial correlation functions of the first order, i.e. $p=1$, for diamond (top) and sawtooth (bottom) chains numerically computed  for different temperatures $k_B T= 10 E_J$ (red), $0.99 E_J$(blue) and $0.01 E_J$ (brown). The frustration $f=0$ was used. }
    \label{fig:FM-correlators}
\end{figure}

\begin{figure}
    \includegraphics[width=0.95\columnwidth]{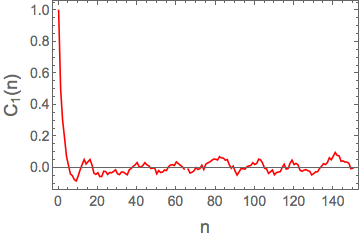}
    \includegraphics[width=0.95\columnwidth]{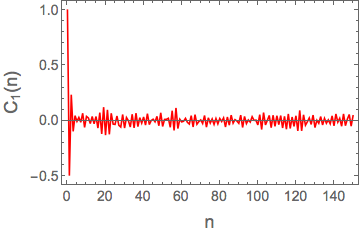}
    \caption{(Color online) The numerically computed spatial correlation functions of the first order, i.e. $p=1$, for diamond (top) and sawtooth (bottom) chains at frustration  $f=0.8$. The temperature $k_B T = 0.01 E_J$ was chosen. At variance with the low  frustration, ferromagnetic regime, the correlation function vanishes beyond few unit cell distances. The observed fluctuations on larger distances are due to finite temperature and finite size effects. }
    \label{fig:HighFrustrationCn}
\end{figure}

The Monte-Carlo results fully support the analytical predictions. Indeed, for both the diamond and sawtooth chains the spatial correlations are clearly ferromagnetic for low frustration $f$ ($f<2/3$ for diamond and $f<3/4$ for sawtooth chains), i.e. the long-range spatial correlations develop in $C_1(n)$ as the temperature is lowered to zero. This behavior is shown in Fig.~\ref{fig:FM-correlators}.  Once the frustration $f$ is large enough, both the diamond and the sawtooth models show spatial correlations in $C_1(n)$  that drop to zero beyond few nearest unit cells (see Fig.~\ref{fig:HighFrustrationCn}). This observation is quantified with the help of the correlation length $\xi_1$ extracted from the fitting  $C_1(n)\propto\exp(-n/\xi_1)$ (see~\eqref{Corrfunction-DC-LT-Nofr} and~\eqref{Corrfunction-ST-LT-Nofr}). As expected $\xi_1$ grows with decreasing temperature for low frustration. However as $f> f^{D(ST)}_c$ , the length $\xi_1$ essentially drops to zero for all temperatures and remains of order one, i.e. of few unit cells  in the highly frustrated phase $f\geq f^{D(ST)}_c$.

%    Figure 7

%    Figure 8
\begin{figure}
    \includegraphics[width=0.95\columnwidth]{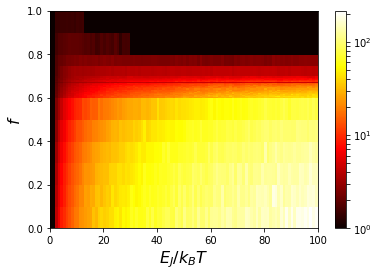}
    \includegraphics[width=0.95\columnwidth]{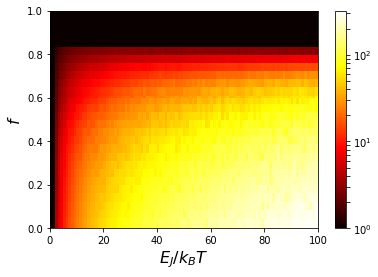}
    \caption{(Color online) The 2d plot showing the dependence of correlation length $\xi_1$ on frustration $f$ and temperature $T$  for diamond (top) and sawtooth (bottom) chains. It was extracted from the Monte-Carlo computation of the spatial correlations functions $C^{D(ST)}_1(n)$.  The length $\xi_1$ decreases for a fixed temperature $T$ as frustration $f$ is increased and essentially it vanishes once the frustration exceeds the critical value $f^{D(ST)}_c$.} 
    \label{fig:xi1TAlpha}
\end{figure}

\begin{figure}
   \includegraphics[width=0.95\columnwidth]{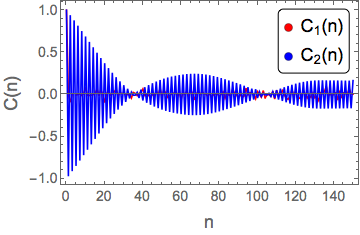}
   \caption{(Color online) The spatial correlation functions for the  diamond chain, $C_p^D(n)$ for $p=1$ (red line) and $p=2$ (blue line) and $f=1$ at $k_B T=0.01 E_J$ . The correlation function with $p=1$  drops to zero beyond the second unit cell, while the $p=2$ correlation function shows long-range features. The fluctuations at large distances are due to stronger fluctuations in $d=1$ systems.}
   \label{fig:DC-Cp12}
\end{figure}

However there are unexpected spatial correlations hidden in the higher order correlation functions $C_p(n)$ with $p>1$.  Indeed, the Monte-Carlo simulation for the diamond chain with the largest frustration $f=1$, shows that the long-range correlations are restored in $C_2^{D}(n)$ for low temperatures. This result  is displayed in Fig.~\ref{fig:DC-Cp12} (similar behavior is observed for the sawtooth chain (not shown)). While $C^{D}_1(n)$ decays to zero beyond $2$ unit cells, the $C_2(n)$ shows long-range oscillatory  behavior with sign alternation. This is in good agreement with the analytical results (see Fig.~\ref{fig:Dchain-Cf}).

%    Figure 9

\begin{figure}
    \includegraphics[width=0.95\columnwidth]{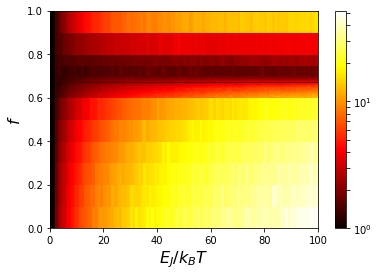}
    \includegraphics[width=0.95\columnwidth]{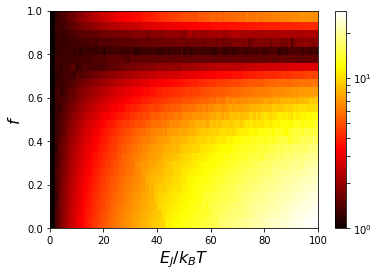}
    \caption{(Color online) The density plot of the correlation length $\xi_2$ (top) and $\xi_3$ (bottom) vs. frustration $f$ and temperature $T$; brighter color corresponds to larger value of the correlation lengths. The data are extracted from the Monte-Carlo simulated $C_2(n)$ and $C_3(n)$ for the diamond and sawtooth chains respectively. For $f < f^{D(ST)}_c$ the $T$-dependence of $\xi_2$, $\xi_3$ follows that of $\xi_1$, reflecting ferromagnetic ordering. For $f >f^{D{ST}}_c$ the long-range order is developing in the higher order spatial correlation functions $C_2$ (diamond) and $C_3$ (sawtooth) as $f\to 1$.}
    \label{fig:xinTAlpha}
\end{figure}

Similarly to the $p=1$ case (see Fig. \ref{fig:xi1TAlpha}) the long-range correlations developing at $f=1$ in the diamond and sawtooth chains can be quantified by the correlation lengths $\xi_2$ and $\xi_3$ ($C_p(n)\propto\exp(-n/\xi_p)$) respectively as shown in Fig.~\ref{fig:xinTAlpha}. In the low frustration phase $f < f^{D(ST)}_c$ both lengths grow with decreasing temperature reflecting the development of the ferromagnetic ordering, similarly to an increase of $\xi_1$. The growth is diminishing as the critical frustration $f^{D(ST)}_c$ value is approached and spatial correlations vanish. Upon entering the highly frustrated regime $f\geq f^{D(ST)}_c$ and approaching $f=1$ the lengths again start to show growth with decreasing temperature and a long-range ordering establishes at $f=1$, $T=0$. Similar effect is observed in higher order correlation functions for several other values of $f > f_c$ for both models.

%    Figure 10

\section{Conclusions}
 
We have studied analytically and using Monte-Carlo simulations the static and dynamic properties of frustrated arrays of Josephson junction classical nonlinear oscillators. Such periodic arrays are characterized by Josephson coupling strengths in every cell that can have different signs. We have considered two models of such frustrated arrays, namely, the diamond and sawtooth chains (see Figs.~\ref{fig:schematic}).
 
In the linear regime the dynamics of these arrays is determined by a multi-band dispersion relation $\omega(k)$ (see Figs.~\ref{fig:DChain} and~\ref{fig:STchain}). Such spectrum can be experimentally accessed through the analysis of the resonant response of the arrays to an applied small amplitude electromagnetic wave. We find that the spectrum of frustrated arrays consists of flatbands for particular values of frustration $f$. Moreover at the critical value of frustration $f_c$ the lowest band becomes flat, and the transition between non-frustrated and frustrated nonlinear regimes occurs at the same value of frustration. We identified the critical values of frustration $f^{D(ST)}_c$ for both diamond and sawtooth chains. 

The transition is reflected in the temperature and frustration dependencies of spatial correlation functions of different orders, $C_p(n)$. In particularly, we obtain that the Josephson junction diamond and sawtooth chains display the ferromagnetic order for low frustration, $f<f_c$, as all Josephson phases (the phase differences between nearest-neighbor vertices) are equal to zero. The ferromagnetic ground state is characterized by the exponential decay of the spatial correlation function $C_p(n)$, where the correlation length $\xi_1$  increases with the decreasing temperature. These results are presented in the top part of Fig.~\ref{fig:Dchain-Cf} (analytical results) and Fig.~\ref{fig:FM-correlators} (Monte-Carlo simulations).

At $f=f_c$ the transition to the frustrated regime occurs. It is characterized by the appearance of the massive degeneracy of the ground state, where any of the two different configurations of the phases, that depend on the frustration value $f$, can be chosen independently for every cell. This results into a drastic suppression of spatial correlations in the correlation function $C_1(n)$ (see the bottom part of Fig.~\ref{fig:Dchain-Cf}(bottom) and Fig.~\ref{fig:STchain-Cf} (analytic results), and Fig.~\ref{fig:HighFrustrationCn} (Monte-Carlo simulations)). The dependence of correlation length $\xi_1$ on temperature and frustration is shown as a density plot in Fig.~\ref{fig:xi1TAlpha}.

However, the long-range correlations are recovered in higher-order correlators $C_p(n)$ with $p>1$ deep in the frustration regime. Indeed, in the limit $f \to 1$ the long-range correlations have been identified for $C^D_2(n)$ for the diamond chain (see Figs.~\ref{fig:Dchain-Cf}(bottom),~\ref{fig:DC-Cp12} and~\ref{fig:xinTAlpha}(top)) and $C^{ST}_3(n)$ for the sawtooth chain (see Fig.~\ref{fig:STchain-Cf} and~\ref{fig:xinTAlpha}(bottom)). Furthermore, we found that the $C^D_2(n)$ changes sign in every cell. The long-range spatial correlations of high-order correlation functions indicate the presence of $2ne$-condensation ($n$ is larger than 1) in the frustrated arrays of Josephson junctions similarly to the model studied in Ref.~\onlinecite{douccot2002pairing} where the long-range correlations have been obtained in $C_2(n)$. 

It is instructive to compare these results to the previously studied models of frustrated arrays: the model elaborated in Ref.~\onlinecite{douccot2002pairing} shows a massively degenerate ground state only for a single value of the frustration, unlike our case, where the degeneracy appears in a whole region of values of frustration $f_c < f < 1$. It is worth noting, that the frustration strength $f=1$ for which $C_p$, $p=2$ and $p=3$ correlators show long-range order for the diamond and sawtooth chains respectively, is not unique. There are other values of frustration $f$ where higher order correlators $C_p(n)$ display long-range behavior at low temperatures. A few examples for the diamond chain are $f=0.75$ with $p=3$, $f\approx 0.71$ with $p=4$, $f=0.81$ with $p=5$, $f\approx 0.85$ with $p=7$ and $f\approx 0.78$ with $p=8$. Interestingly there are no oscillations in $C_5(n)$ for $f=0.81$.
The geometry of the diamond and sawtooth chains studied strongly suggests that similar results hold for other corner sharing chains. The flatband construction of Ref.~\onlinecite{morales2016simple} relying on repetition of mini-arrays appears particularly promising.

Finally, we note that quantum tunneling between two equivalent states in every cell can result in a unique ground state and a corresponding crossover between classical highly degenerated ground state and macroscopic quantum state at low temperatures. 

%\alexei
%Finally it is worth mentioning, that similar results are expected for other frustrated lattices: stub, cross-stitch, $1D$ Lieb, etc.
%}

%\alexei{ALEXEI: I think we should add comparison to previous results: Ref.~\onlinecite{douccot2002pairing} as well as experimental realisations, that you did mention in the discussions. I think we need a comment on the other frustrated lattices - we picked diamond and sawtooth, but there are many more. Also I believe we never give the construction of the groundstates in frustrated regime? Is that necessary?}

\begin{acknowledgments}
  The authors would like to thank S. Flach for useful discussions.  M. V. F. acknowledges the financial support of the Ministry of Education and Science of the Russian Federation in the framework of Increase Competitiveness Program of NUST "MISiS" $K2-2017-067$ and the State Program 3.3360.2017. This work was also supported by the Institute for Basic Science in Korea (IBS-R024-D1).
\end{acknowledgments}

\bibliography{general,flatband,josephson}

\end{document}